# SEPARATION OF BIOCRUDE PRODUCED FROM HYDROTHERMAL LIQUEFACTION OF FAECAL SLUDGE WITHOUT ANY SOLVENT

**H M Fairooz Adnan\*, Md Khalekuzzaman\*, Md. Atik Fayshal\* and Md. Mehedi Hasan\*\***

*\*Department of Civil Engineering, Khulna University of Engineering & Technology (KUET), Khulna-9203, Bangladesh*

*\*\*Institute of Disaster Management, Khulna University of Engineering & Technology (KUET), Khulna, 9203, Bangladesh*

## ABSTRACT

*Production of high-quality biocrude using Hydrothermal Liquefaction (HTL) process from various waste-stream gaining more concern recently due to fossil fuel scarcity around the world. Faecal sludge (FS) significantly pollutes the environment and emits greenhouse gases (GHGs) has gained more concern recently. A more effective "waste to energy" strategy would use Hydrothermal Liquefaction (HTL) of FS to create high-quality biocrude. In this study faecal sludge is used as raw biomass due to its abundance, low cost, and easy availability. After HTL operation, product separation is getting challenging. Current developed studies observed the separation of aqueous and biocrude oil products occurs during the HTL process more popularly with the use of an organic solvent which is quite expensive. Focusing on this critical issue, this study aims to separate the biocrude and aqueous phase without using any solvent by gravity separation technique. From FTIR analysis data it showed that centrifuged at 6000 rpm partial separation of biocrude and aqueous phase (AP) was noticed. however, at 9000 rpm, FTIR analysis showed that biocrude samples included aliphatic hydrocarbons, phenols, and esters where no signs of any carbon chain were found at AP which indicated the products are successfully separated. The separated Crude portion had the higher A-Factor (0.68) and lower C-Factor (0.58) value which indicates the oil quality was immature grade of lower kerogen type II (i.e., moderate oil-prone). This low-cost technique can be economically advantageous for commercial-scale biocrude production.*

## INTRODUCTION

The rapid rate of global growth causes high levels of energy usage and environmental pollution, which in turn contributes to considerable greenhouse gas (GHG) emissions, climate change, and global warming. Global warming is mostly caused by emissions from fossil fuels, according to the Intergovernmental Panel on Climate Change (IPCC). 89% of the world's CO2 emissions in 2018 came from industry and fossil fuels (*World Energy Outlook 2021*, n.d.). The use of fossil fuels to produce energy is not environmentally friendly or sustainable (Gani, 2021). The process of creating energy from waste is one of the sustainable energy creation strategies. In addition to reducing GHG emissions and addressing the global warming catastrophe, the "Waste to Energy" technology method also establishes a sustainable path for energy generation (Iqbal & Kang, 2021).

Faecal Sludge (FS) is a growing environmental concern worldwide due to rapid population growth. Without effective management, the globe produces between 250 and 300 million metric tons of FS annually, which is a significant cause of soil, surface water, and groundwater contamination (Cairns-Smith et al., 2014). FS management is a significant concern in developing countries that hinders them from meeting sustainable development goals (SDGs). Gasification is a simple process for converting solid waste into energy or power since they have limited liquid content (Belgiorno et al., 2003). FS is not suitable for this approach due to its high moisture content (around 90%). Through a thermochemical conversion like hydrothermal liquefaction, FS is easily transformed into biocrude (HTL). HTL is a promising technique that uses water as a catalyst to convert wet biomass into biofuel at high temperatures (180-375 °C) and high pressure (4-30 MPa) (Huang et al., 2019). Recent studies, based on the assumption of a higher heating value (HHV), proposed FS as a highly prospective biomass for the HTL process (Kabir et al., 2022). Huang et al., (2019) suggest that sludge (activated, sewage, etc.) hydrothermally liquefied in the presence of the co-substate can produce biocrude with excellent yield and quality.

Since dichloromethane (DCM) recovers biocrude with the best yield/energy efficiency and has the lowest energy consumption ratio, compared to other solvents like acetone or toluene, the majority of the research were carried out in batch HTL trials (Watson et al., 2019). The result of the HTL process is combined with DCM, and the aqueous phase, biocrude, and biochar are separated (Hossain et al., 2022). The presence of an organic solvent, however, has reportedly been shown to significantly alter the characteristics of biocrude. However, the current research were mostly conducted at the batch experiment level, and as a result, their HTL behavior and subsequent AP characteristics may vary due to changes, such as stirring type, heat transfer, and mass transfer in batch and continuous systems. To provide relevant and realistic data for the development of scale-up processes, the impact of organic solvent on product yields has previously been documented. After all, utilizing this organic solvent in scaled-up production appears to be unfavorable from an economic perspective as well as environmentally hazardous (Guo et al., 2022).

The recovery of AP from a continuous HTL process for FS may further increase the process efficiency and is receiving growing interest. The goal of our work is to promote a more economical management of FS with hydrothermal liquefaction as the core technology and get a deeper understanding of this process chain. In this work, two consecutive steps have been investigated along one processing route: first to separate the AP from the product of HTL process without solvent and to determine the product maturity while separating without solvent. Different centrifugal speeds were applied and the composition of the AP and biocrude have been examined. After a purification treatment, the obtained AP could be further evaluated for irrigation or other purposes. In the upcoming years, this cutting-edge solid-partially decomposed wastes co-liquefaction method may serve as a dual-purpose solution for combating both climate change and global warming as well as sustainable biofuel generation.

## MATERIALS AND METHODS

**Biomass Sampling and Preparation**

Faecal sludge was collected from the septic tank in the residential portion of the Khulna University of Engineering & Technology (KUET) campus. The FS sample was then immediately kept overnight at 4°C. After being homogenized, the raw FS was kept for this investigation. In Table 1, the proximal components of FS are shown.

**Table 1:** Proximate analysis of fecal sludge

| Components | Weight % |
|---|---|
| Moisture Content (MC) | 88.7 ± 0.6 |
| Total Solids (TS) | 11.3 ± 0.5 |
| Volatile Solids (VS) | 7.2 ± 0.4 |
| Ash Content (AC) | 2.3 ± 0.1 |
| Fixed Carbon (FC) | 1.8 ± 0.2 |

Fixed carbon (FC), % = 100 – sum of (VS%, MC%, AC%).

**Hydrothermal Liquefaction (HTL) Process and Product Separation**

In a stainless steel (SS 304) 25 mL batch reactor with a working capacity of 10 mL, HTL tests were carried out. The biomass was directly put into the reactor during each HTL operation, which was manually sealed with a copper gasket and capped. The Carbolite ESF 12/2, Bamford, Sheffield model furnace was then used to heat it up. The HTL experiment was carried out at 300°C with a heating rate of 65°C/min and a reaction duration of 60 min (Dandamudi et al., 2020). The compressed gases were released at the time the head was opened. Previous studies used organic solvent in the process of product separation. (Hossain et al., 2022)) used dichloromethane (DCM) solution in product separation. But there was no solvent utilized in this investigation. The end product of the HTL operation was put into 5 mL polypropylene centrifuge tubes. The product was then centrifuged for 5 min using a mrc CENHBR-600ML tabletop centrifuge to separate the aqueous phase from the combination of biocrude, biochar, and aqueous phase. Two specific centrifugal speed was used during this separation process. Khalekuzzaman et al. (2021), (Xu et al., 2019) use 4000 rpm for 10 minutes to separate each layer with solvent. Guo et al. (2022) applied centrifugation of 8000 rpm for 5 minutes without using solvent. In this study, centrifugation of 6000 rpm for 5 minutes and 9000 rpm for 5 minutes were applied to separate aqueous phase. The HTL products were divided into two layers: the top layer, the aqueous phase, and the biochar-biocrude/oil layers remaining in bottom layer. Subsequently, the aqueous phase and biochar-biocrude layer were separated using a 5 mL syringe. Finally, the aqueous phase was separated from biochar and biocrude. The experimental flow diagram is presented in Figure 1.

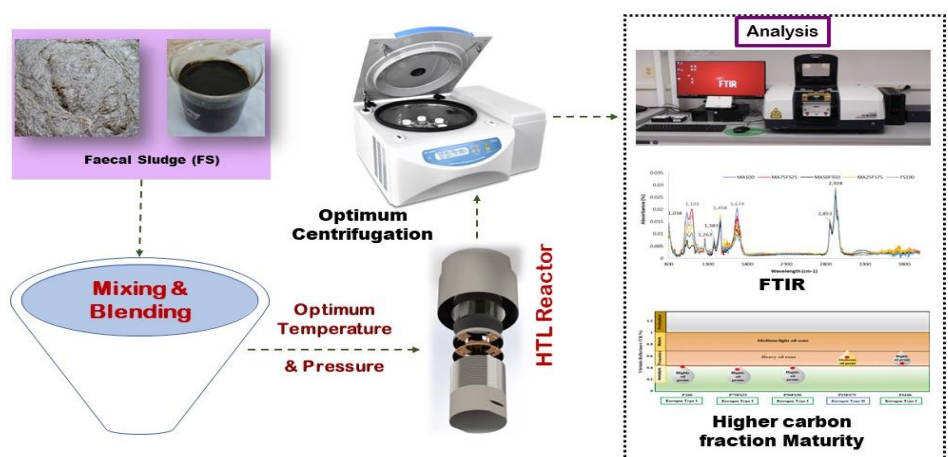

Figure 1: Experimental flow diagram

**Analytical Methods**

FTIR analyses were conducted on the biocrude-biochar, and aqueous samples of FS to determine vibrations, chemical compounds, and functional groups after separation. Shimadzu's (IRTracer-100) spectrophotometer was used to perform FTIR analysis at room temperature on spectra with resolutions of 2 cm$^{-1}$ and ranging from 800 cm$^{-1}$ to 4000 cm$^{-1}$. The transmittance unit was used to gather the IR data, which was afterwards converted to absorbance (%) and analyzed using OriginPro 2018 and MS Excel.

The Ganz and Kalkreuth approach was used to analyze the A-Factor (aliphatic/aromatic bands) and C-Factor (carbonyl/aromatic bands) from the FTIR spectra of biocrude samples from FS (Ganz & Kalkreuth, 1991). Using OriginPro 2018, the peak heights of 2930 cm$^{-1}$, 2860 cm$^{-1}$, 1705 cm$^{-1}$, and 1630 cm$^{-1}$ were chosen for this study in terms of absorbance (%) (Adedosu & Sonibare, 2005). Equations (1) and (2) were used to derive the A-factor and C-factor. The A-Factor and C-Factor analyses were used to determine the kerogen type (oil intensity) and maturity index of the biocrude samples, respectively. (Islam et al., 2022).

$$\text{A-Factor} = (2930 \text{ cm}^{-1} + 2860 \text{ cm}^{-1}) / (2930 \text{ cm}^{-1} + 2860 \text{ cm}^{-1} + 1630 \text{ cm}^{-1}) \quad \text{......................(1)}$$

$$\text{C-Factor} = 1705 \text{ cm}^{-1} / (1705 \text{ cm}^{-1} + 1630 \text{ cm}^{-1}) \quad \text{...............................................(2)}$$

The biocrude oil source type/parent kerogen type and thermal maturity were classified using a vitrinite reflectance equivalent grid diagram based on the A-Factor vs. C-Factor. Type I of the four kerogen types is thought to be the most oil-prone. Type II denotes crude with a moderate oil tendency, Type III denotes crude with a low oil tendency, and Type IV denotes crude with no possibility for the presence of petroleum hydrocarbons (Kareem et al., 2016). In addition, the maturity levels of biocrude were classified based on vitrinite reflectance (VR %). The biocrude thermal maturation parameter (VR%) was divided into four categories: postmature (thermogenic/dry gas zone), peak mature (light oil zone), and immature (biogenic gas zone) (Hossain et al., 2022).

**RESULTS AND DISCUSSION**

FTIR spectroscopy was used to examine the presence of functional groups and the mode of vibration in biocrude, biochar, and aqueous samples. After centrifugation at 6000 rpm, FTIR spectroscopy was analyzed for both the biocrude-biochar component and the aqueous phase. The findings demonstrated that the biocrude ratios had nearly identical patterns of spectra (functional groups), confirming that the biocrude samples had identical chemical structures. The presence of long-chain aliphatic hydrocarbons was indicated by the broad and strong peak in the biocrude samples, which was discovered in the ranges of 2800 cm$^{-1}$ - 3000 cm$^{-1}$ (-CH$_3$) and 1350 cm$^{-1}$ - 1460 cm$^{-1}$ (-CH$_2$) (Li et al., 2018). However, the band between 1590 and 1800 cm$^{-1}$ (C=O) showed that biocrude had ketones, carboxylic acids, aldehydes, and esters due to the hydrolysis of cellulose and hemicellulose (Feng et al., 2018). In addition, the absorbance peaks in the 1024 cm$^{-1}$ to 1100 cm$^{-1}$ (C-O) region showed that biocrude included phenolic and alcohol components (Zhang et al., 2018). These are formed as a result of the hydrolysis or depolymerization of the lignin and carbohydrates found in biomass cells.

The FTIR spectra also revealed that aqueous samples contained aliphatic hydrocarbon and methyl (-CH3, -CH2-), carbonyl (C= O, O – H), and ether (C – O) functional groups. The presence of these groups indicated the aqueous phase also have carbon chain which cannot be acceptable because of the loss of potential carbon chain.

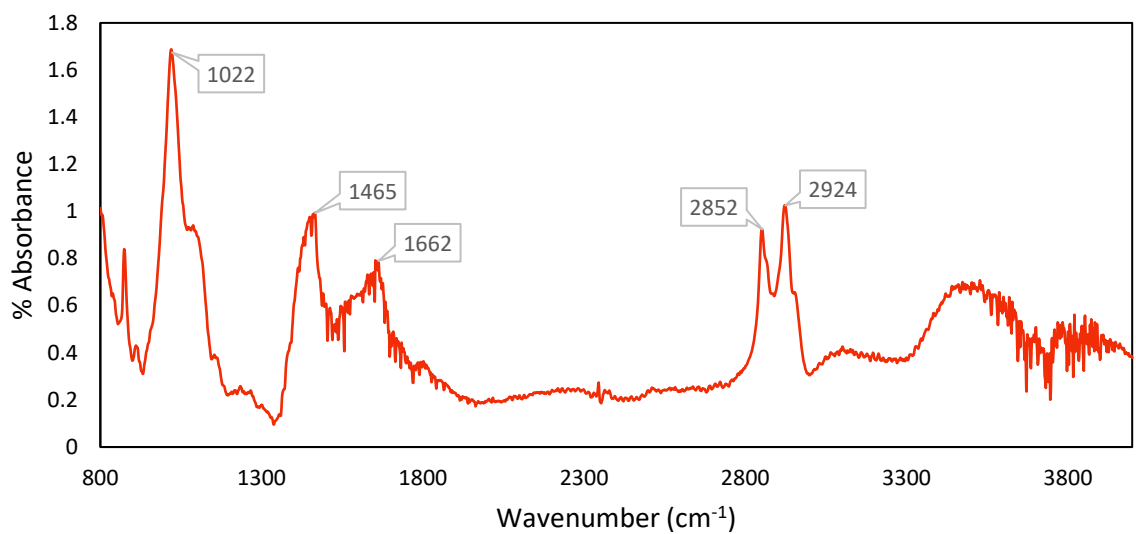

(a)

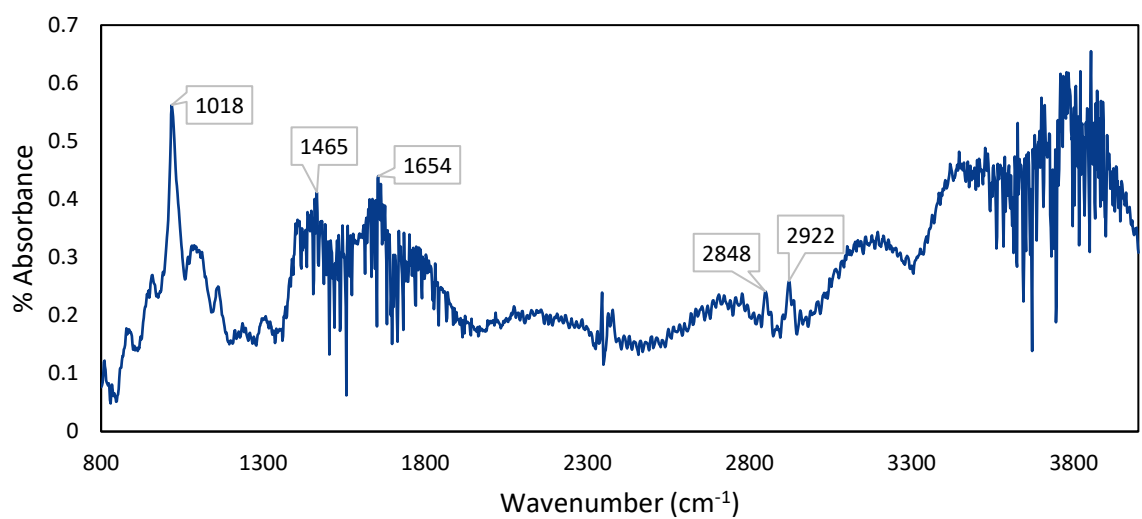

(b)

Figure 2: FTIR spectra of (a) biocrude-biochar and (b) aqueous phase after centrifugation of 6000 rpm

After Centrifugation of 9000 rpm FTIR spectroscopy were also analyzed for both biocrude-biochar portion and aqueous phase. In biochar-biocrude phase, there also sharp and strong peak at the range of 2800 $cm^{-1}$ – 3000 $cm^{-1}$ and 1350 $cm^{-1}$ – 1460 $cm^{-1}$ which indicates the presence of long-chain aliphatic hydrocarbons. On the other hand, strong peak at the range of 1590 $cm^{-1}$ to 1800 $cm^{-1}$ suggested the presence of presence of ketones, carboxylic acids, aldehydes, and esters in biocrude through the hydrolysis of cellulose and hemicellulose (C=O) (Feng et al., 2018). Besides, for the depolymerization or hydrolysis reaction mechanism of carbohydrate and lignin cell structure of biomass, the absorbance peaks shown in the range of 1024 $cm^{-1}$ – 1100 $cm^{-1}$ indicated the presence of alcohol (C-O) and phenolic (C=O) compounds in biocrude-biochar sample.

In the aqueous sample, there only two peaks at 1645 $cm^{-1}$ and 3385 $cm^{-1}$. There is no peak at the range of 2800 $cm^{-1}$ - 3000 $cm^{-1}$ and 1350 $cm^{-1}$ – 1460 $cm^{-1}$. Which indicates there is no functional group of aliphatic hydrocarbon and methyl (-$CH_3$, -$CH_2$-), carbonyl (C= O, O – H), or ether (C – O). The absence of these functional group indicates that there is no loss of Carbon chain. IR spectra of the samples show broad bands of the stretching of an NH$_2$ group at 3385 and 1645 cm$^{-1}$ (Qiu & Gao, 2003).

The maturity of biocrude oil is affected by a number of variables, including the A-factor, C-factor, and VR%. On the basis of FTIR analysis, the A-factor and C-factor were determined and presented in Table 2. Higher A-Factor and lower C-Factor indicated the presence of higher aliphatic bands and lower carbonyl bands in biocrude. Results revealed that the sample had the A-Factor (0.68) and C Factor (0.58). C-factor also plays a major role

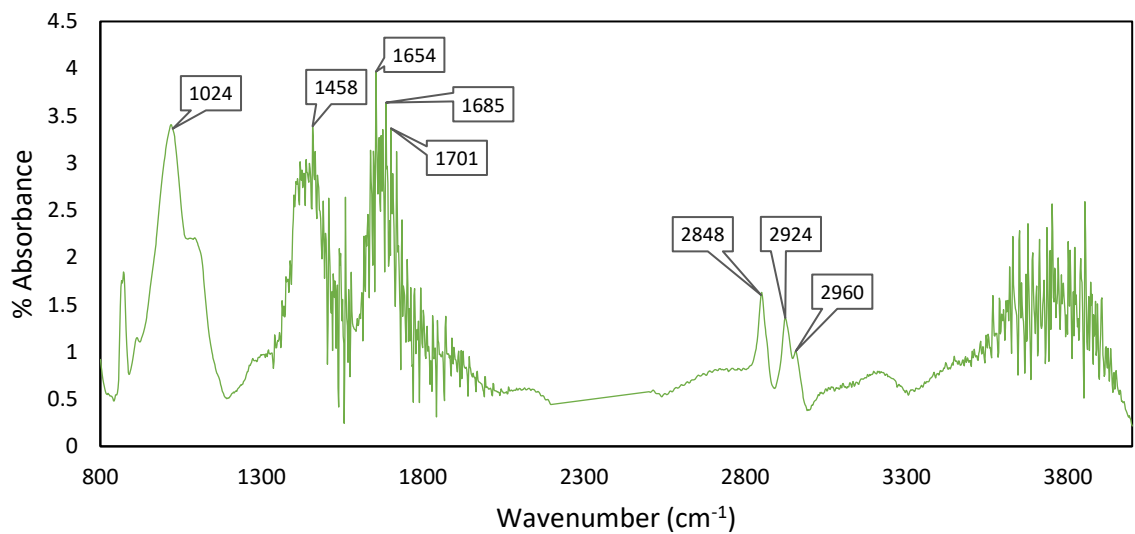

(a)

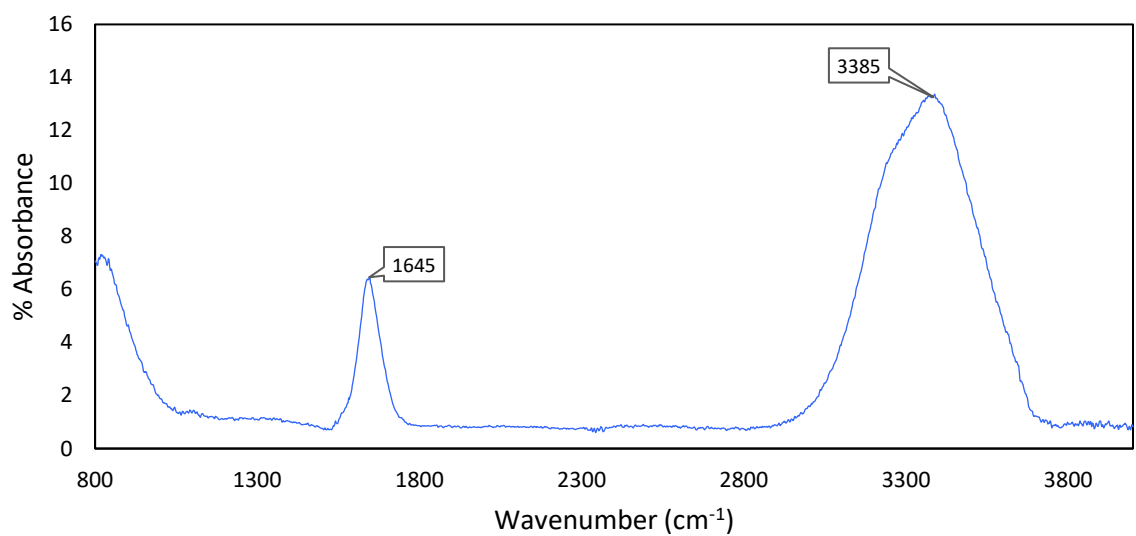

(b)

Figure 3: FTIR spectra of (a) biocrude-biochar and (b) aqueous phase after centrifugation of 9000 rpm

In dictating the VR (%) value. When C-Factor decreases, the VR (%) value increases, and thus the biocrude quality also improves. Higher A-Factor and lower C-factor also improve the oil-prone nature of biocrude. Biocrude maturity level was determined based on VR (%) and presented in Table 2. The VR (%) for the biocrude samples was 0.32. According to this VR (%) range, the biocrude samples were of the immature type. Biocrude source rock type (kerogen type) revealed that the samples were kerogen type-II that represented moderately oil-prone crude.

Table 2: Maturity and oil intensity parameters of biocrude

| Parameters | Value |
|---|---|
| A-Factor | 0.68 |
| C-Factor | 0.58 |
| Vitrinite Reflectance (VR%) | 0.32 |
| Maturity Level | Immature |
| Kerogen Type | Type-II |
| Oil Intensity | Moderate Oil-Prone |

## CONCLUSIONS

In this study, Faecal Sludge was processed via HTL in a batch reactor at certain temperatures, and HTL liquid products were separated without solvent. Centrifugation of 9000 rpm showed the perfect separation of the aqueous phase. The FTIR study showed the presence of aliphatic hydrocarbon, ester, phenolic, and alcohol components in biocrude-biochar. While there was no carbon functional group present in the aqueous sample. FTIR data indicated the product after separation had higher A-factor and lower C-factor value which results improvement of oil quality. However, biocrude maturity found immature with kerogen type II (moderate oil-prone). However, one of the major obstacles in crude separation process was addition of solvent. Solvents are very costly and further solvent extraction is also difficult and time consuming. Focusing this issue, the study demonstrated the separation technique of crude portion without using any organic solvent. Separation process without using solvent would be cost effective approach to extract crude from the product of HTL process. Additionally, since organic solvent is not included in the HTL process output, energy consumption will be minimized as well as significant capital cost can be saved at pilot and commercial scale crude production. This novel approach of crude separation without solvent might be a sustainable low-cost approach for energy-dense light biocrude oil production which can prove effective during this fuel crisis across the entire world.